\documentclass[aps,prc,groupedaddress,showpacs,twocolumn,floatfix]{revtex4}

\usepackage{graphicx}
\usepackage{longtable}

\def\deg{^\circ}
\def\gtorder{\mathrel{\raise.3ex\hbox{$>$}\mkern-14mu
 \lower0.6ex\hbox{$\sim$}}}
\def\ltorder{\mathrel{\raise.3ex\hbox{$<$}\mkern-14mu
 \lower0.6ex\hbox{$\sim$}}}
\def\mugegm{\mu_p G_E / G_M}
\def\gegm{G_E / G_M}
\def\ge{G_E}
\def\gm{G_M}
\def\gep{G_E}
\def\gmp{G_M}
\def\sigr{\sigma_R}

\begin{document}

\title{Global analysis of proton elastic form factor data with two-photon
exchange corrections}

\author{J. Arrington}
\affiliation{Argonne National Laboratory, Argonne, Illinois 60439, USA}

\author{W. Melnitchouk}
\affiliation{Jefferson Lab, Newport News, Virginia 23606, USA}

\author{J. A. Tjon}
\affiliation{Physics Department, University of Utrecht, The Netherlands}

\date{\today}

\begin{abstract}

We use the world's data on elastic electron--proton scattering and
calculations of two-photon exchange effects to extract corrected values
of the proton's electric and magnetic form factors over the full $Q^2$
range of the existing data.
Our analysis combines the corrected Rosenbluth cross section and
polarization transfer data, and is the first extraction of $\gep$ and
$\gmp$ including explicit two-photon exchange corrections and their
associated uncertainties.
In addition, we examine the angular dependence of the corrected cross
sections to look for possible nonlinearities of the cross section
as a function of $\varepsilon$.

\end{abstract}
\pacs{25.30.Bf, 13.40.Gp, 14.20.Dh}

\maketitle

\section{Introduction}\label{sec:intro}

As one of the most fundamental observables which characterize the
composite nature of the nucleon, electromagnetic form factors have over
the past few decades provided considerable insight into the nucleon's
internal structure~\cite{hydewright04, gao05, perdrisat06, arrington07a}.
In the commonly used one-photon exchange approximation to
electron--nucleon scattering, the form factors depend only on the
four-momentum transfer squared, $-Q^2$.
At low $Q^2$ the Fourier transforms of the form factors give information
on the charge and magnetization distributions of the nucleon.

The expectation from perturbative QCD is that the $Q^2$ dependence of the
Sachs electric, $\ge(Q^2)$, and magnetic, $\gm(Q^2)$, form factors should
be the same at large $Q^2$, and early data indeed suggested approximate
scaling of the ratio $\gegm$ with $Q^2$~\cite{walker94}. These data were
analyzed using the Rosenbluth or longitudinal-transverse (LT) separation
technique, in which the form factor ratio is extracted by examining the
elastic cross section as a function of the scattering angle, $\theta$. The
cross section at $\theta=180\deg$ depends only on the magnetic form
factor, while the cross section at smaller angles is a combination of
magnetic and electric contributions.  At large $Q^2$, the contribution
from the electric form factors is small, and here the technique has
reduced sensitivity to $\ge$.

Recent experiments at Jefferson Lab~\cite{punjabi05, gayou01, gayou02},
utilizing the polarization transfer (PT) technique to measure the ratio
$\gegm$, found the surprising result that $\ge$ decreases more rapidly
than $\gm$ at large $Q^2$. This indicates a substantially different
spatial distribution of the charge and magnetization of the
nucleon~\cite{kelly02}, raising questions as to the impact of angular
momentum and relativistic effects on the nucleon structure~\cite{lu99,
buchmann01, miller03, matevosyan05, gross06, kvinikhidze06, wang07}.  The
virtue of the PT technique is that it is sensitive only to the ratio
$\gegm$, rather than the sum of a large magnetic and small electric
contribution, and so does not suffer from the dramatically reduced
sensitivity to $\ge$ of the LT separation at large $Q^2$. However, the two
techniques disagree significantly even in the region where they both yield
precise results~\cite{arrington03a}, and a puzzle has existed about the
origin of the discrepancy.

An accurate determination of the proton electric and magnetic form factors
is also important as these impact on our knowledge of other quantities
whose extraction is sensitive to $\gep$ and $\gmp$.
In some cases, uncertainties in the proton form factors will be magnified,
yielding even larger corrections in other observables.

A number of recent theoretical studies of two-photon
exchange (TPE) in elastic $ep$ scattering have been performed
\cite{guichon03, blunden03, chen04, afanasev05a, blunden05a, kondratyuk05,
jain06, borisyuk06a} (see also Ref.~\cite{carlson07} for a recent review).
These indicate that TPE effects give rise to a strong angular-dependent
correction to the elastic cross section, which can lead to large
corrections to the LT-extracted $\gegm$ ratio.  In fact, the results of
quantitative calculations based both on hadronic intermediate
states~\cite{blunden05a, kondratyuk05, jain06, borisyuk06a, kondratyuk07}
and on generalized parton distributions~\cite{chen04, afanasev05a},
provide strong evidence that TPE effects can account for most of the
difference between the LT and PT data sets.

While the TPE analyses are very suggestive of the resolution of the
problem, there exists some residual discrepancy between the TPE-corrected
data sets.  Moreover, the effects of the TPE correction on the extraction
of $\ge$ and $\gm$ have in practice been estimated using a linear
approximation to the correction over an arbitrary angular range
\cite{blunden03, blunden05a}, which can yield significant uncertainties
in evaluating the effect of TPE on the form factor extractions,
especially at high $Q^2$.  An accurate determination of $\ge$ and $\gm$
requires analyzing the elastic scattering data taking into account the
TPE effects {\em directly} on the cross sections.  This is the primary
aim of the present paper.

Our approach will be to apply the TPE corrections directly to the cross
section data, estimate the uncertainties in the TPE corrections, and
compare TPE-corrected results to polarization measurements. We then use
the corrected cross section results, combined with polarization transfer
measurements, to extract the best possible values and uncertainties for $\ge$
and $\gm$ of the proton.

The outline of this paper is as follows.  In Sec.~\ref{sec:formalism} we
review the formalism of elastic $ep$ scattering in the presence of TPE
effects and discuss the TPE corrections which we apply to the data. We
also study the effect of TPE on the ratio of electron--proton to
positron--proton elastic cross sections.  In Sec.~\ref{sec:global}, TPE
corrections are applied to the cross section and polarization transfer
measurements, and a combined analysis of all measurements is performed.
We provide parameterizations of the corrected form factors, as well as a
direct fit to the elastic cross section, without correction for TPE
effects. Finally, in Sec.~\ref{sec:conclusion} we summarize our results
and discuss future work.

\section{Elastic $e p$ Scattering}\label{sec:formalism}

\subsection{Born approximation}\label{sec:born}

In the one-photon exchange or Born approximation the differential cross
section is given in terms of the Born amplitude ${\cal M}_0$ by:
\begin{equation}
{ d\sigma_0 \over d\Omega }
= \left( {\alpha \over 4 M Q^2 }{E' \over E} \right)^2
  \left| {\cal M}_0 \right|^2\
= \sigma_{\rm Mott}\frac{1}{\varepsilon (1+\tau)} \sigr\ ,
\label{eq:sigma0}
\end{equation}
where $\sigma_{\rm Mott}$ is the cross section for scattering on a
point particle, $E$ and $E'$ are initial and final electron energies,
respectively, $\tau = Q^2/4M^2$, and $\alpha = e^2/4\pi$ is the
electromagnetic fine structure constant.  In performing a Rosenbluth
separation, it is convenient to work with the reduced cross section,
$\sigma_R$:
\begin{equation}
\sigr = \tau  \gm^2(Q^2) + \varepsilon  \ge^2(Q^2)\ ,
\label{eq:sigmaR}
\end{equation}
where $\varepsilon = \left[ 1 + 2 (1+\tau) \tan^2{(\theta/2)} \right]^{-1}$
is the virtual photon polarization parameter.  This decomposition allows
a direct separation of $\ge$ and $\gm$ from the $\varepsilon$ dependence
of $\sigr$.  In the Born approximation the form factors are functions of
only a single variable, $Q^2$. Effects beyond one-photon exchange
introduce additional $\varepsilon$ dependence.

The polarization transfer method involves the scattering of a polarized
electron beam from an unpolarized target, with measurement of the
polarization of the recoiling proton.  In the Born approximation the
ratio of the transverse to longitudinal recoil polarizations
yields~\cite{akhiezer74}:
\begin{equation}
R = \mu_p { \ge \over \gm }
  = -\mu_p \sqrt{\tau (1+\varepsilon)\over 2 \varepsilon}
	{P_T \over P_L }\ ,
\label{eq:poltrans}
\end{equation}
where $\mu_p$ is the proton magnetic moment, and $P_T$ ($P_L$) is the
polarization of the recoil proton transverse (longitudinal) to the proton
momentum in the scattering plane.  It should be noted that both $P_T$ and
$P_L$ individually depend only on the ratio $\gegm$, with $P_T$ ($P_L$)
being strongly (weakly) dependent on $\gegm$.  In most experiments,
however, the ratio of the two polarization components is used, with
$P_T$ providing sensitivity to the form factor ratio, and the ratio to
$P_L$ yielding a result that is independent of the overall beam and
target polarization and/or analyzing power of the detector.  One has
similar sensitivity to $\gegm$ in measurements of the beam-target
asymmetry~\cite{dombey69, alguard76, donnelly86}.

\subsection{TPE: Unpolarized $e p$ scattering}\label{sec:tpe_unpolarized}

With the inclusion of radiative corrections to order $\alpha$, the
elastic scattering cross section in Eq.~(\ref{eq:sigma0}) is modified
according to:
\begin{equation}
{d\sigma_0 \over d\Omega} \to  {d\sigma \over d\Omega} =
{d\sigma_0 \over d\Omega}\ (1 + \delta_{\rm RC})\ ,
\end{equation}
where $\delta_{\rm RC}$ represents one-loop corrections, including
vacuum polarization, electron and proton vertex, and two-photon
exchange corrections, in addition to inelastic bremsstrahlung for
real photon emission~\cite{mo69}.

As discussed in Refs.~\cite{blunden03, blunden05a}, the amplitude for the
one-loop virtual corrections, ${\cal M}_1$, can be written as the sum of
a factorizable term, proportional to the Born amplitude ${\cal M}_0$, and
a non-factorizable component, $\overline{\cal M}_1$:
\begin{equation}
{\cal M}_1 = f(Q^2,\varepsilon) {\cal M}_0 + \overline{\cal M}_1\ .
\label{eq:m1}
\end{equation}
The ratio of the full, ${\cal O}(\alpha)$ cross section to the
Born cross section can then be written as:
\begin{equation}
1 + \delta_{\rm RC} =
{ \left| {\cal M}_0 + {\cal M}_1 \right|^2 \over \left| {\cal M}_0 \right|^2 },
\end{equation}
with the correction $\delta_{\rm RC}$ given by:
\begin{equation}
\delta_{\rm RC} = 2 f(Q^2,\varepsilon)
       + { 2 \Re\{{\cal M}_0^\dagger \overline{\cal M}_1\}
	  \over |{\cal M}_0|^2}.
\label{eq:delta}
\end{equation}

The contributions to the functions $f(Q^2,\varepsilon)$ from the electron
vertex, vacuum polarization, and proton vertex terms depend only on $Q^2$,
and hence do not affect the LT separation, aside from an overall
normalization factor.  Of the factorizable terms, only the IR-divergent
TPE correction contributes to the $\varepsilon$ dependence of the virtual
photon corrections~\cite{blunden03}.

The non-factorizable terms contained in $\overline{\cal M}_1$ arise
from the finite nucleon vertex and TPE corrections, which depend 
explicitly on nucleon structure.  For the proton vertex correction,
the $\varepsilon$ dependence is weak, and will not significantly
affect the LT analysis~\cite{maximon00}.  For the inelastic
bremsstrahlung cross section, the amplitude for real photon
emission can also be written in the form of Eq.~(\ref{eq:m1}).

In the soft photon approximation, in which one of the two exchanged 
photons is taken to be on-shell, the full amplitude is completely 
factorizable.  A significant $\varepsilon$ dependence arises due
to the frame dependence of the angular distribution of the emitted
photon.  These corrections, together with external bremsstrahlung,
contain the main $\varepsilon$ dependence of the radiative corrections,
and are accounted for in standard experimental analyses.  They are
generally well understood, and in fact enter differently depending
Âon whether the electron or proton are detected in the final  
state~\cite{afanasev01, qattan05}.

The only remaining term at ${\cal O}(\alpha)$ that can introduce a
non-negligible $\varepsilon$ dependence, namely the non-factorizable
part of the TPE contribution, is typically not accounted for in cross
section extractions.  Its effects can be included by considering the
interference of the total TPE and Born amplitudes, according to:
\begin{equation}
\delta_{2\gamma} =
{ 2 \Re \left\{ {\cal M}_0^\dagger\ {\cal M}_{2\gamma} \right\}
  \over \left| {\cal M}_0 \right|^2}\ ,
\label{eq:delta_eff}
\end{equation}
where the TPE amplitude ${\cal M}_{2\gamma}$ in principle includes all
possible (off-shell) hadronic intermediate states.

In typical experimental analyses of electromagnetic form factor data,
radiative corrections are implemented using a formalism based on the
Mo \& Tsai (MT) prescription~\cite{mo69,tsai71}.  While the detailed
implementation has been improved over the years~\cite{walker94,ent01},
the treatment of TPE has remained unchanged from the original work.
The TPE effects are partially included in the MT approach by
approximating the TPE amplitude by its infrared (IR) divergent part.

A more detailed examination of the loop integrals~\cite{maximon00}
yields the same IR-divergent contribution as Mo \& Tsai~\cite{mo69},
$\delta_{\rm IR}({\rm MT})$, but also yields an IR-finite contribution.
The IR-finite part, which is usually neglected in the standard data
analyses, was found in Refs.~\cite{blunden03, blunden05a} to have a
significant $\varepsilon$ dependence, and is explicitly included here.
To isolate the effect of the additional TPE contributions on the data,
Blunden et al.~\cite{blunden03, blunden05a} consider the difference:
\begin{equation}
\Delta  \equiv  \delta_{2\gamma} - \delta_{\rm IR}({\rm MT})\ ,
\label{eq:Delta_dif}
\end{equation}
in which the IR divergences cancel.  We will follow this approach and
therefore reference the $\varepsilon$ dependence of the full calculation
of $\delta_{2\gamma}$ with that of $\delta_{\rm IR}({\rm MT})$.

The results for the difference, $\Delta(\varepsilon,Q^2)$, between the
full calculation and the MT approximation are most significant at low
$\varepsilon$, and essentially vanish at large $\varepsilon$.  At the
lower $Q^2$ values, $\Delta$ is approximately linear in $\varepsilon$,
but significant deviations from linearity are observed with increasing
$Q^2$, especially at smaller $\varepsilon$~\cite{blunden05a}.

The effect of TPE on the LT extractions of $\ge$ is greatest at large
$Q^2$ since the cross section here is not very sensitive to
$\ge$.  However, the TPE corrections to the cross section have a
relatively weak $Q^2$ dependence, and are therefore still significant
at \textit{low} $Q^2$ values, where many high-precision
measurements of nucleon or nuclear structure are performed that
require accurate knowledge of the proton form factors.

\subsection{TPE: Theoretical uncertainty}\label{sec:uncertainty}

For the dominant nucleon elastic TPE contribution, the main uncertainty
arises from the input form factors at the internal $\gamma^* NN$
vertices in the loop diagrams.  Because the intermediate state nucleon
is off-shell, in principle the ``half off-shell'' form factors here need
not be the same as the free nucleon form factors, and can explicitly
depend on the intermediate nucleon four-momentum.  The off-shell
dependence is of course unknown, but in practice it is sufficient, at
least at low $Q^2$, to approximate these by the free form factors.

For the calculation in Ref.~\cite{blunden05a} it was necessary for
technical reasons to parameterize the internal form factors by sums of
monopole functions, which were fitted to realistic parameterizations of
form factor data.  In the actual analysis~\cite{blunden05a} the form
factor parameterization was taken from the global fit in
Ref.~\cite{mergell96}.  For comparison, more recent fits to
data~\cite{brash02,arrington04a} were also used, and the sensitivity of
the TPE correction $\Delta(\varepsilon,Q^2)$ to the particular form was
found to be negligible up to $Q^2 \sim 10$~GeV$^2$.

In the present analysis we use the more recent global form factor 
parameterization from Ref.~\cite{arrington04a} with $\gep$ constrained
by the PT data.  A three-monopole fit to this parameterization provides
a very good fit for $\gep$ and $\gmp$ over the range where both the
form factors are constrained by data.  Above $Q^2=6$~GeV$^2$, the
three-monopole fit to $\gep$ becomes unconstrained, leading to unstable
results for the TPE corrections.  Here we instead use a dipole
approximation for the form factors, which guarantees a smooth 
extrapolation to high $Q^2$, and matches onto the asymptotic high-$Q^2$
behavior expected from perturbative QCD.  The difference between the
TPE results at $Q^2=6$~GeV$^2$ with the ``realistic'' and dipole form
factors is not significant, which is consistent with the findings in
Ref.~\cite{blunden05a}.  Similar calculations from Ref.~\cite{borisyuk06a}
find results that are in very good agreement with those of
Ref.~\cite{blunden05a}.

Estimates of the contributions to $\Delta(\varepsilon,Q^2)$ from
higher-mass intermediate states have been made by a number of authors
\cite{drell57, drell59, greenhut69, kondratyuk05, tjon07, kondratyuk07}.
In a recent analysis, Kondratyuk et al.~\cite{kondratyuk05} evaluated the
contribution of the $\Delta(1232)$ resonance, which is known to play an
important role in hadronic structure, between $Q^2 = 1$ and 6~GeV$^2$.
The $\Delta(1232)$ intermediate state contribution was found to be
smaller in magnitude than the nucleon contribution, with an opposite
sign at backward scattering angles where the TPE effects are largest.
For realistic choices of the $\gamma N\Delta$ vertex, and couplings
obtained from a hadronic model analysis of Compton scattering off the
nucleon~\cite{kondratyuk01}, the magnitude of the $\Delta(1232)$
contribution was found to be around 1/4 of that of the nucleon at low
and intermediate $Q^2$.  The $\Delta(1232)$ therefore tends to cancel
some of the TPE effect from the nucleon elastic intermediate state,
modifying the cross section by between $-1\%$ and 2\%.
At larger $Q^2$ the magnitude of the $\Delta(1232)$ contribution
increases, along with that of the nucleon elastic, although above
$Q^2 \sim 10$~GeV$^2$ the reliability of the calculation is more
questionable.

Effects of higher mass, spin- and isospin-1/2 and 3/2 resonances were
also estimated recently by Kondratyuk \& Blunden~\cite{kondratyuk07}.
The excited states included in that analysis were the $P_{11}$ $N(1440)$
Roper resonance, the $D_{13}$ $N(1520)$, and the odd-parity $S_{11}$
$N(1535)$ for isospin-1/2, and the $S_{31}$ $\Delta(1620)$ and $D_{33}$
$\Delta(1700)$ isospin-3/2 resonances.  The photocouplings for these
states were taken from the model~\cite{kondratyuk01} of nucleon Compton
scattering at energies up to the first and second resonance regions,
and a universal dipole form factor was applied for each of the
$\gamma N \to$\ resonance transitions.  In addition, the assumption was
made that the transitions are mostly magnetic, with the electric being
much smaller, and the Coulomb couplings taken to be zero.

The overall scale of the higher mass contributions is approximately an
order of magnitude smaller than the nucleon and $\Delta(1232)$ for
$Q^2 \ltorder 6$~GeV$^2$.  Moreover, these alternate in sign, with the
$P_{11}$, $D_{13}$ and $D_{33}$ contributions being mostly negative
(as the nucleon), while the $S_{11}$ and $S_{31}$ mostly positive.
For larger $Q^2$ this model, as with all models based on hadronic
degrees of freedom, is unlikely to be reliable.

Clearly the inclusion of the higher mass hadronic intermediate states is
considerably more model dependent than the nucleon elastic or even the
$\Delta(1232)$ contribution, in view of uncertainties in the various
photocouplings, the transition form factors, and the efficacy of
representing high energy off-shell nucleon excitations by states of
zero width.  Effects of overlapping resonances, as well as the
multihadron continuum, or non-resonant background, are expected to be
increasingly important at larger $Q^2$.  One could in principle constrain
the high mass spectrum phenomenologically by Compton scattering data at
high $Q^2$.  Unfortunately these are not yet available, and one must rely
on theoretical guidance.

From the point of view of quark-hadron duality, one expects that at
large $Q^2$ the intermediate state spectrum can be saturated either by
including a large set of hadronic resonances, or by summing over quark
degrees of freedom~\cite{melnitchouk05}.  Although the approaches are
in principle complementary, if $Q^2$ is large enough a quark level
calculation may provide a more efficient description.  Such calculations
have been performed in the context of generalized parton distributions
(GPDs)~\cite{chen04, afanasev05a} within the ``handbag'' approximation,
in which the scattering of the virtual photon takes place off a single
quark in the proton.  This should be the dominant (leading twist)
contribution at high $Q^2$, although at intermediate $Q^2$, processes
where the virtual photon interacts with different quarks in the proton
(higher twist) are still important.

One of the uncertainties in the calculation is the choice of GPD,
and in Ref.~\cite{afanasev05a} a Gaussian valence ansatz is used.
The calculation has a small but noticeable difference when using their
modified Regge GPD, and has a similar dependence on the assumed quark
mass.  In addition, for the leading twist approximation to be
applicable, both $Q^2$ and the invariant mass squared $s$ of the
intermediate state are required to be large.  In practice the
constraint $Q^2, s > M^2$ is used, which restricts the kinematic
reach of the calculation to moderate and large $\varepsilon$.  For
$Q^2 \sim 2-6$~GeV$^2$, one is restricted to $\varepsilon > 0.2-0.4$.
The results show significant $\varepsilon$ dependence with clear
deviations from linearity at intermediate and large $\varepsilon$,
and a weak $Q^2$ dependence.  The overall trend of the $\varepsilon$
dependence is similar to that in Ref.~\cite{blunden05a}, but smaller
in magnitude.  At high $Q^2$ there appears little $\varepsilon$
dependence for $\varepsilon > 0.5$, which yields little modification
to the LT extraction of $\gep/\gmp$ dominated by data at large
$\varepsilon$.  At the highest $Q^2$ points the partonic contribution
explains about 1/2 of the discrepancy between the LT and PT results.

The indications from both the hadronic and partonic higher mass
calculations are therefore that there is an additional effect at the
level of a few percent on top of the nucleon elastic contribution,
albeit with a sizable theoretical uncertainty.  Given the difficulty
in obtaining a reliable quantitative estimate of the higher mass
contributions to TPE at high $Q^2$, rather than rely on a specific model,
we take a more phenomenological approach.

At low $Q^2$, the correction is approximately linear with $\varepsilon$.
While the calculations suggest increasing nonlinearities, they differ
in the detailed $\varepsilon$ dependence, and the data are consistent with
very small nonlinearities~\cite{tvaskis06}.  We assume therefore a
linear $\varepsilon$ dependence for the form of the extra TPE correction.
We make an estimate of the $Q^2$ dependence of the additional terms
from the calculations of the higher resonance states~\cite{kondratyuk07}
and the GPD-based model~\cite{afanasev05a}, and apply this as the
additional TPE correction.  For $Q^2 < 1$~GeV$^2$, we do not apply any
additional TPE effect, as all indications are that it is very
small.  For $Q^2 > 1$~GeV$^2$, we take a $Q^2$ dependence that grows
slowly with $Q^2$:
\begin{equation}
\delta_{2\gamma}^{*} = 0.01 ~ [\varepsilon-1] ~ \frac{\ln{Q^2}} {\ln{2.2}}
~~~~~~~(Q^2>1~{\rm GeV}^2),
\label{eq:extraTPE}
\end{equation}
with $Q^2$ in GeV$^2$, and apply a 100\% uncertainty.  This yields a
correction linear in $\varepsilon$ that decreases the $\varepsilon=0$
cross section by 1\% at $Q^2=2.2$~GeV$^2$, by 2\% at 4.8~GeV$^2$,
and by 3\% at 10.6~GeV$^2$.  This correction is then added to the
calculated $\delta_{2\gamma}$.

As a further check, we compare the extraction of $\gegm$ from the
TPE-corrected cross section data and the polarization transfer
measurements, which is discussed in Sec.~\ref{sec:rosenbluth} and
shown in the bottom panel of Fig.~\ref{fig:corrected_lt}, below.
The additional correction brings the high-$Q^2$ LT $\gep/\gmp$ points
into good agreement with the PT data, although the error on the
corrected LT data is considerably larger at high $Q^2$.

While the effect of the TPE corrections on $\gmp$ is smaller, it can
potentially have a large impact on global fitting.  For instance,
the TPE corrections for $\gep$ shift around 10 -- 15 data points by
$1 - 2 \sigma$, but for $\gmp$ there are of order 30 -- 40 points
affected by a $2 - 3 \sigma$ shift.  In addition, even though the TPE
corrections to $\gmp$ are less dramatic than those for $\gep$, the
uncertainties in the TPE effects when extracting $\gmp$ at large $Q^2$
will still be important.

\subsection{TPE: $e^- p$ vs. $e^+ p$}\label{sec:tpe_positron}

While TPE corrections modify the unpolarized cross section, it is 
difficult to isolate TPE corrections from the Born cross section
experimentally without independent knowledge of the proton form factors.
Only deviations from the linear $\varepsilon$ dependence predicted by
Eq.~(\ref{eq:sigmaR}) can be observed, and in practice these are found
to be very small~\cite{tvaskis06}.  However, \textit{direct} experimental
evidence for the contribution of TPE can be obtained by examining the
ratio of $e^+p$ and $e^-p$ cross sections.
The Born amplitude changes sign under the interchange
$e^- \leftrightarrow e^+$, while the TPE amplitude does not.
The interference therefore has the opposite sign for electron and
positron scattering, which will show up in the experimental ratio
$R^{e^+e^-} = \sigma_{e^+ p}/\sigma_{e^- p}
	 \approx (1 - 2 \Delta$),
where $\Delta$ is defined in Eq.~(\ref{eq:Delta_dif}).
Interference between electron and proton bremsstrahlung also yields
a small difference, but this is corrected for in the extraction of
$R^{e^+e^-}$.

Although the current data on elastic $e^- p$ and $e^+ p$ scattering
are sparse, there are some experimental constraints from early data
taken at SLAC~\cite{browman65,mar68}, Cornell~\cite{anderson66}, 
DESY~\cite{bartel67} and Orsay~\cite{bouquet68} (see also
Ref.~\cite{arrington04b} and references therein).  The data are
predominantly at low $Q^2$ and at forward scattering angles,
corresponding to large $\varepsilon$ ($\varepsilon \agt 0.7$), where
the calculated 2$\gamma$ exchange contribution is small ($\alt 1\%$).
Nevertheless, the overall trend in the data reveals a small enhancement
in $R^{e^+e^-}$ at the lower $\varepsilon$ values~\cite{arrington04b}.

\begin{figure}[ht]
\includegraphics[width=8.4cm]{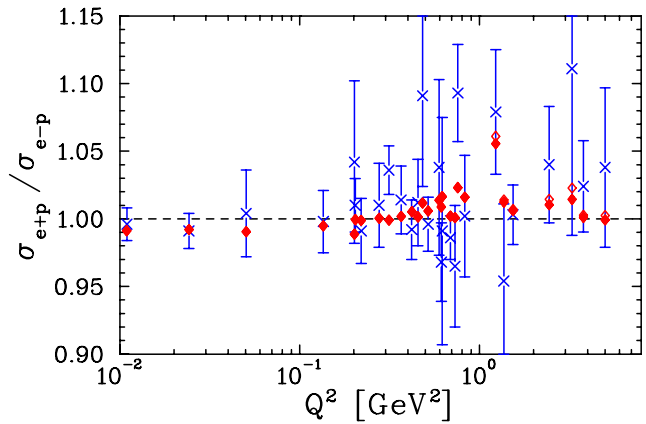}
\caption{(Color Online) Ratio of elastic $e^+ p$ to $e^-p$ cross sections.
	The data (crosses) are taken from~Ref.~\cite{arrington04b}
	and references therein.  The filled diamonds show the ratio
	using the TPE calculation of Ref.~\cite{blunden05a}, while
	the open diamonds include the additional contribution in
	Eq.~(\ref{eq:extraTPE}), which modifies the high-$Q^2$
	correction at low $\varepsilon$.
\label{fig:Ree}}
\end{figure}

Figure~\ref{fig:Ree} shows the extracted ratio, $R^{e^+e^-}$, compared
with the calculation of Ref.~\cite{blunden05a} (filled diamonds), and
with the additional high mass contribution in Eq.~(\ref{eq:extraTPE})
(open diamonds).  The calculation is in good agreement with the data,
although the errors on the data points are quite large.  Clearly better
quality data at backward angles and moderate $Q^2$, where an enhancement
of up to $\sim$10\% is predicted, would be needed for a more definitive
test of the TPE mechanism.  There is a planned experiment to perform a
precise ($\sim$1\%) comparison of $e^- p$ and $e^+ p$ scattering at
$Q^2=1.6$~GeV$^2$ and $\varepsilon \approx 0.4$ at the VEPP-3 storage
ring~\cite{vepp_proposal}.  An experiment to provide broader
$\varepsilon$ and $Q^2$ coverage~\cite{e04116} at Jefferson Lab is
approved to make such measurements up to $Q^2 = 2 - 3$~GeV$^2$ using
a beam of $e^+ e^-$ pairs produced from a secondary photon beam.

\subsection{TPE: Polarization observables}\label{sec:tpe_polarization}

Naively, one expects the corrections to the spin-dependent cross
sections to be of the same order of magnitude ($\ltorder 5$\%) as the
corrections to the unpolarized cross sections.  For unpolarized
scattering, this yields a large correction to the extracted value of
$\ge$ at large $Q^2$, where the total contribution from $\ge$ is also
small.  The polarization measurements are directly sensitive to the
ratio $\mugegm$, and so the TPE corrections to this ratio will be of
the same magnitude as the correction to the cross section; the large
magnification of the effect observed in the Rosenbluth separations does
not occur for polarization measurements.

Both the hadronic~\cite{blunden05a} and partonic~\cite{afanasev05a}
approaches can be used to estimate the TPE corrections to the recoil
polarization measurements.  While the calculations are not strictly
valid in the same kinematical regimes, both predict the corrections
to be extremely small, $\ltorder 1\%$, except for large $Q^2$ and low
$\varepsilon$ values.  The corrections to the existing data are less
than 2\% for all settings except for the largest $Q^2$ point, where the
corrections are of order 5\%, which is still much smaller than the
statistical uncertainty of the measurement.  However, the sign of the
correction is different in the two approaches.  As the sign of this
correction is not known empirically, and all indications are that it is
very small compared to the experimental uncertainties, we do not apply
a TPE correction to the polarization measurements.
An experiment~\cite{e04019} at Jefferson Lab will map out the
$\varepsilon$ dependence of the polarization transfer at fixed $Q^2$,
which will provide the first direct information on the TPE correction
to polarization transfer.

\section{Global Analysis}\label{sec:global}

In this section we perform a combined analysis of cross section and
polarization data, including both the standard radiative corrections
{\em and} the TPE contributions described above.  In
Ref.~\cite{arrington07b}, the form factors were extracted after
applying Coulomb distortion corrections, but only cross section
data were included, and the analysis was limited to very low $Q^2$
values.  This is the first global analysis of form factors in which TPE
effects are included in a consistent way from the outset.  We apply the
TPE corrections to the data, and repeat the global analysis of
Ref.~\cite{arrington04a}, with a few important modifications:

\begin{itemize}

\item
We include additional low $Q^2$ cross section data~\cite{frerejacque66,
ganichot72, dudelzakphd}, more recent cross section results
\cite{christy04, qattan05}, and the full set of polarization
transfer and beam-target asymmetry measurements~\cite{milbrath99,
pospischil01, gayou01, gayou02, strauch02, punjabi05, maclachlan06,
jones06, crawford07, ron07}.

\item
We use TPE corrections following the formalism of Blunden et
al.~\cite{blunden05a}, rather than the phenomenological correction used in
Ref.~\cite{arrington04a}. At high $Q^2$ (above 1~GeV$^2$) we supplement this
with an additional TPE correction (Eq.~(\ref{eq:extraTPE})), with a 100\%
uncertainty, to estimate the contributions from the high mass intermediate
states and related (conservative) uncertainties.

\item
We use a different fitting function, based on Ref.~\cite{kelly02}, and
include high-$Q^2$ constraint points in the fit.  The goal is to have a
fit valid at both very low $Q^2$ and the highest $Q^2$ values of the
existing data, with a sensible extrapolation to higher $Q^2$ values.

\end{itemize}

In addition to extracting parameterizations of $\ge$ and $\gm$ from the
global fit, we also take the corrected cross section data in small $Q^2$
bins, and perform direct LT separations.  This allows a test of the
consistency between the TPE-corrected cross section results and the
polarization measurements.  Using the same $Q^2$ bins we also perform
a combined fit to the $\varepsilon$ dependence of the cross sections
and the polarization data, to extract both values and uncertainties for
$\ge$ and $\gm$.

Finally, we provide a parameterization of the TPE-{\it uncorrected}
cross section measurements.  This parameterization is nearly identical
to taking the TPE-corrected form factors and applying the TPE calculation
used here, but does not require an explicit calculation of the TPE
effects.

\subsection{Global fitting procedure and results}\label{sec:globfit}

The combined analysis uses the same general approach outlined in
Refs.~\cite{arrington03a, arrington04a}.  As in the previous analyses,
we remove the small angle data from Ref.~\cite{walker94}, and apply
updated radiative correction factors for lepton loop diagrams to some
of the older experiments which did not include these corrections.
In Ref.~\cite{qattan05} a high precision measurement of the elastic
cross section was aimed specifically at the ratio $\mugegm$.  Their
quoted systematic uncertainty is separated into three components:
normalization, point-to-point, and a linear $\varepsilon$ dependent
correction which impacts $\mugegm$, but does not yield random
fluctuations or nonlinearity relative to the linear fit.
We use only normalization and point-to-point uncertainties, and to
include these data in the fit we increased the point-to-point
systematic uncertainty from 0.45\% to 0.5--0.55\% (depending on $Q^2$)
such that the uncertainty on $\mugegm$ using the inflated point-to-point
uncertainty matched the uncertainty as extracted in the more detailed
analysis.

After applying the TPE correction to the raw cross sections, we perform
a global fit to all of the cross section, polarization transfer, and
beam-target asymmetry measurements using the form:
\begin{equation}
\ge, \gm / \mu_p
= \frac{1+\sum_{i=1}^n a_i \tau^i}{1+\sum_{i=1}^{n+2} b_i \tau^i}\ .
\label{eq:fitfcn}
\end{equation}
We take $n=3$, which yields 8 fit parameters for each form factor,
along with a normalization factor for each independent set of cross
section measurements.  The fit includes 569 cross section and 54
polarization transfer data points, and yields a reduced $\chi^2$ of
0.770.  Taking $n=2$ yields similar fits, with a reduced $\chi^2$ of
0.802, and changes that are below the uncertainties in the form factors;
$\gm$ changes by $\ltorder$1\% up to $Q^2=10$~GeV$^2$, $\ge$ by less
than 3\% up to $Q^2=5$~GeV$^2$.  The normalization factors are
consistent between the $n=2$ and $n=3$ fits, with an RMS variation
below 0.2\%.  The polarization transfer data yield a contribution to
$\chi^2$ of 52 for the 54 data points, indicating good consistency with
the cross section data. The parameters from the combined fit are given in
Table~\ref{tab:combined_fit}, and shown as solid lines in
Fig.~\ref{fig:fitanddata}.  The parameterization is valid up to at least
$Q^2=6$~GeV$^2$ for $\ge$, and $Q^2=30$~GeV$^2$ for $\gm$.

\begin{table}[ht]
\caption{Fit parameters for the extracted proton electric and magnetic
	form factors, using the parametrization of Eq.~(\ref{eq:fitfcn}).
\label{tab:combined_fit}}
\begin{ruledtabular}
\begin{tabular}{crr}
Parameter & $\gm/\mu_p$	& $\ge$~~ \\
\hline
 $a_1$  &  --1.465~~ &    3.439~~ \\
 $a_2$  &    1.260~~ &  --1.602~~ \\
 $a_3$  &    0.262~~ &    0.068~~ \\ \hline
 $b_1$  &    9.627~~ &   15.055~~ \\
 $b_2$  &    0.000~~ &   48.061~~ \\
 $b_3$  &    0.000~~ &   99.304~~ \\
 $b_4$  &   11.179~~ &    0.012~~ \\
 $b_5$  &   13.245~~ &    8.650~~ \\
\end{tabular}
\end{ruledtabular}
\end{table}

The fit function in Eq.~(\ref{eq:fitfcn}) is chosen as it gives
reasonable behavior in both the limits of low $Q^2$ and high $Q^2$.
We constrained the parameters in the denominator ($b_i$ in
Eq.~(\ref{eq:fitfcn})) to be positive, to avoid fits where both the
numerator and denominator pass through zero at the same $Q^2$,
yielding narrow divergences in the fit.
Finally, extra high $Q^2$ ``data'' points were added to the fit
to prevent uncontrolled behavior in the large $Q^2$ limit.
The high $Q^2$ constraint points were $\gm = 0.7\ G_D$, where
$G_D = [1+Q^2/(0.71\ {\rm GeV}^2)]^{-2}$ is the dipole form factor,
at $Q^2 = 50, 100, 200$, and 400~GeV$^2$, and $\ge=0$
at $Q^2 = 10, 15, 20$, and 25 GeV$^2$.
In both cases, the uncertainties were taken to be 50\%, 100\%, 150\%,
and 200\% of $G_D$, respectively.  This has a negligible effect in the range of
the data, but prevents ``extreme'' behavior in the region where the data do not
provide a meaningful separation between $\ge$ and $\gm$.

\subsection{Rosenbluth analysis}\label{sec:rosenbluth}

The combined analysis yields a fit for $\ge$ and $\gm$, as well as
normalization factors for the various cross section measurements.
However, it does not give a clear indication of the consistency between
the cross section and polarization data.  In this section, we use the
normalization factors from the combined global fit to normalize the
individual cross section data sets, and perform a global analysis using
only the cross section data binned in $Q^2$.  The results from the
TPE-corrected cross section analysis can then be compared directly to
the PT measurements, to see to what extent the TPE corrections resolve the
discrepancy.

\begin{figure}[ht]
\includegraphics[width=8.4cm,angle=0]{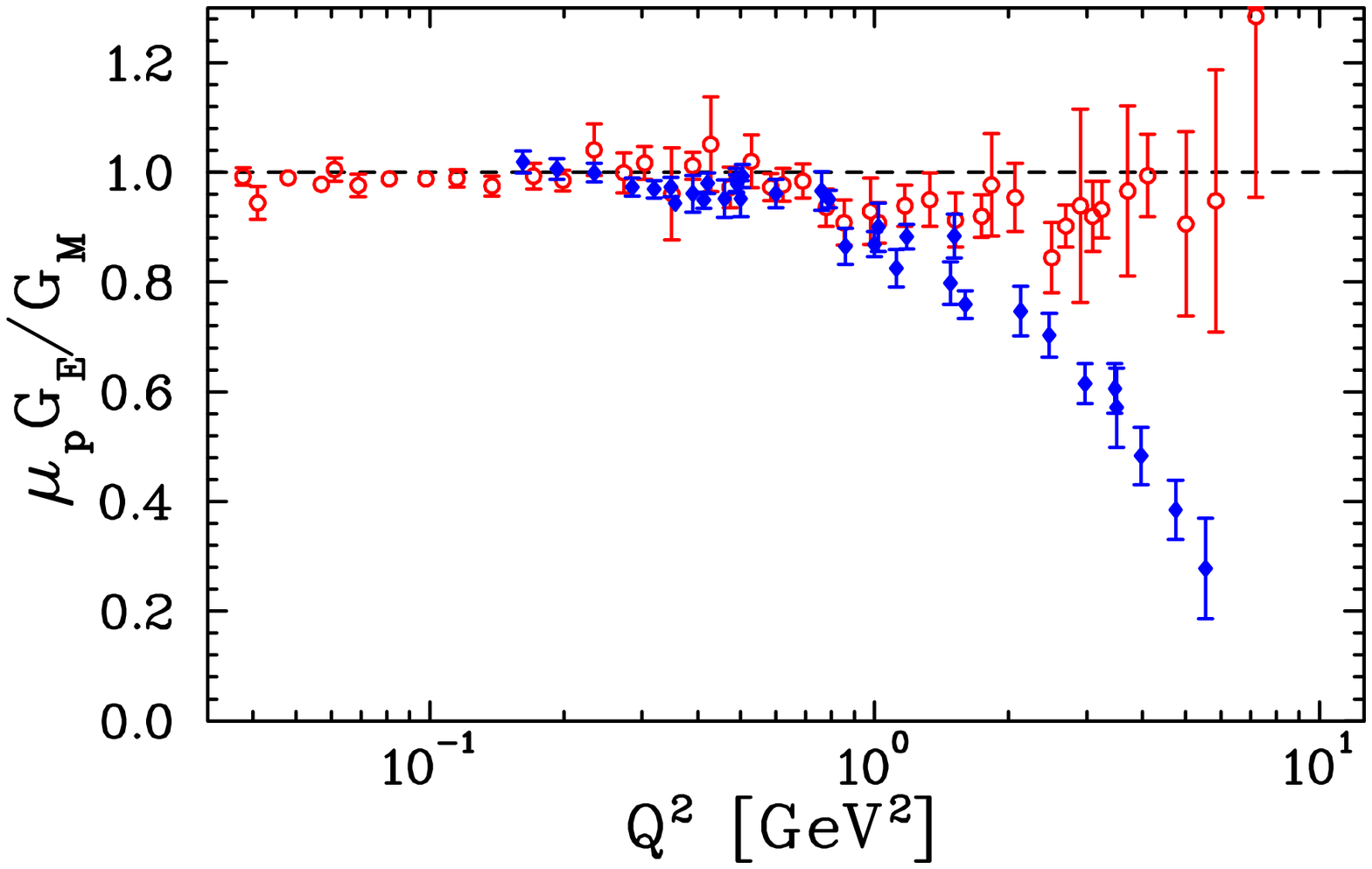}
\includegraphics[width=8.4cm,angle=0]{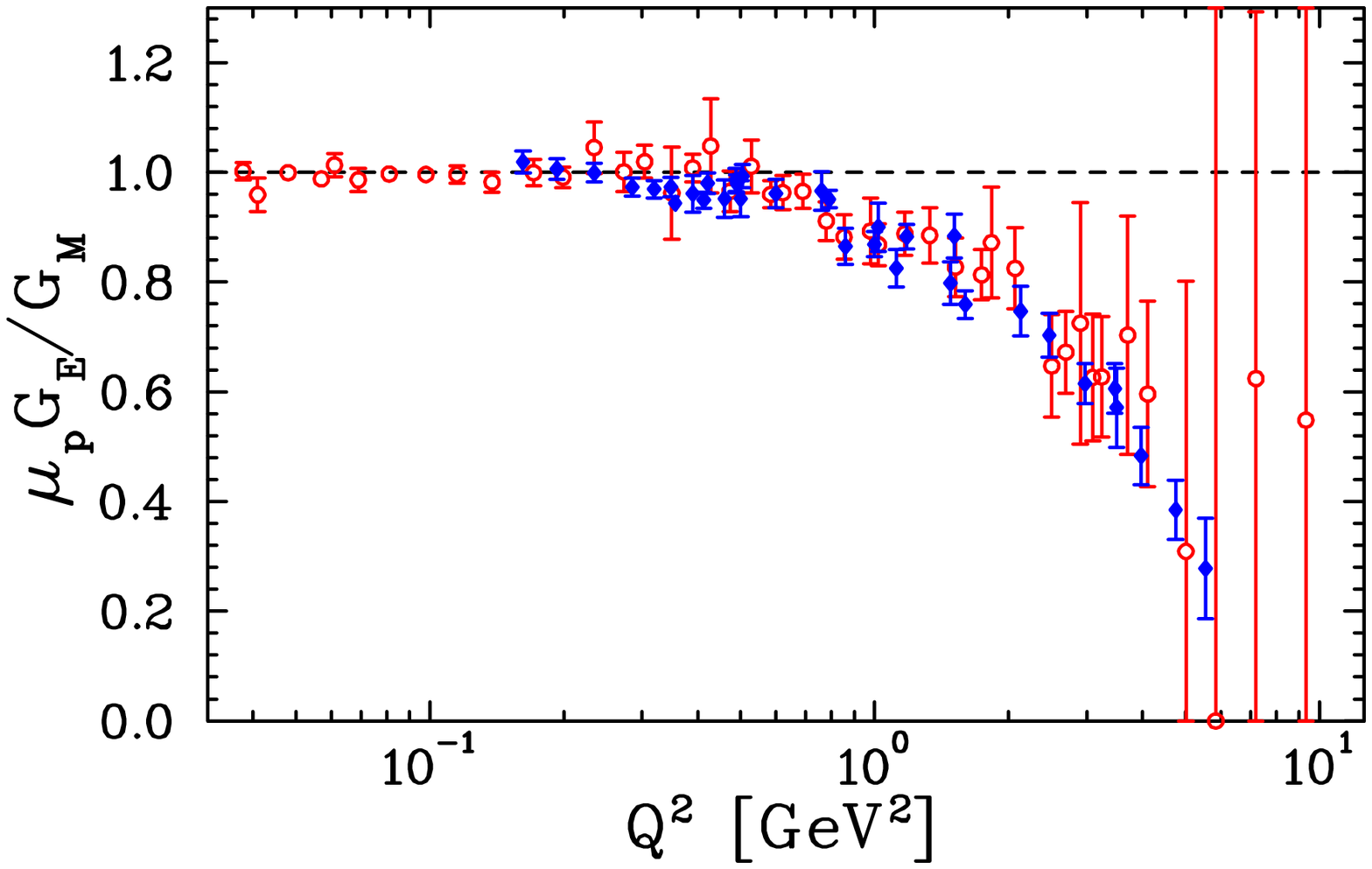}
\caption{(Color Online) Ratio $\mugegm$ extracted from polarization
	transfer (filled diamonds) and LT measurements (open circles).
	The top (bottom) figure shows LT separations without (with) TPE
	corrections applied to the cross sections.  
\label{fig:corrected_lt}}
\end{figure}

We begin by taking cross section data in small $Q^2$ bins and interpolating
each cross section to the mean $Q^2$ value using the fit from
Tab.~\ref{tab:combined_fit}.  The cross section is known well enough that
the uncertainty associated with the interpolation is small. The bin size is
chosen to keep the interpolation correction small, typically at the 1\%
level, and an additional uncertainty equal to 3\% of the correction is
applied to the corrected point.  We then perform an LT separation
to extract $\gegm$.  We determine a systematic uncertainty associated
with the scale factors by taking each data set and varying its
normalization by the estimated scale uncertainty.  Note that
because we have used the global fit to estimate the normalization
factors, we have a smaller scale uncertainty than
quoted for the individual measurements.  We take the scale
uncertainty after the global fit to be the smaller of 1\% or half of
the originally quoted normalization uncertainty.  For the analysis including TPE
corrections, we also take the additional TPE correction arising from
the high mass intermediate states and vary it by $\pm 100\%$ to estimate
the additional TPE uncertainty for large $Q^2$.

In Fig.~\ref{fig:corrected_lt} we show a comparison of the PT data
(filled diamonds) and the LT separation data (open circles).  The data
sets are completely independent, except for the use of the normalization
factors from the global fit of Sec.~\ref{sec:globfit}.  The top panel
shows the global LT analysis without applying TPE corrections, while
the bottom is with TPE corrections. The uncertainties in the LT
separation increase significantly at large $Q^2$ values when the TPE
corrections are applied.  This is in part due to the uncertainty assigned to
the high $Q^2$ TPE corrections, but mainly due to the fact that the LT
separation is a measure of $(\gegm)^2$ rather than $\gegm$, so a reduction
in the slope yields a constant absolute uncertainty in $(\gegm)^2$, which
maps into a larger absolute uncertainty in $\gegm$.

Note that for very low $Q^2$, the TPE correction does not go to zero, as
was also pointed out in Ref.~\cite{borisyuk06b}.
In fact, the correction to the unpolarized cross section becomes zero
for $Q^2 \approx 0.3$~GeV$^2$, and changes sign for lower $Q^2$ values.
This appears to be a largely model-independent result, which persists
even for the case of point-like nucleons. Thus, even the extremely low $Q^2$
extractions are modified by TPE. With the TPE correction applied, the
average value of $\mugegm$ for $Q^2<0.2$~GeV$^2$ goes from 0.988(4) to
0.997(4), showing that $\ge$ and $\gm$ have consistent low-$Q^2$ behavior.

One of the unique features of the TPE correction is that it introduces
nonlinearity in the $\varepsilon$ dependence of the cross section.
Although part of the TPE contribution is linear in $\varepsilon$,
observation of nonlinearity in the data would provide direct evidence
of TPE effects in elastic scattering.  To quantify the amount of
nonlinearity in the data, we fit the reduced cross section to a
quadratic in $\varepsilon$, as in Ref.~\cite{tvaskis06}:
\begin{equation}
\sigma_R = P_0 [ 1 + P_1 (\varepsilon-0.5) + P_2 (\varepsilon-0.5)^2 ]\ ,
\label{eq:quadfit}
\end{equation}
where $P_2$ represents the fractional $\varepsilon$-curvature parameter,
relative to the average ($\varepsilon=0.5$) reduced cross section.
With the inclusion of TPE corrections, the average nonlinearity
parameter, $\langle P_2 \rangle$, is found to increase from
$1.9 \pm 2.7\%$ to $4.3 \pm 2.8\%$.  While the extracted nonlinearity
increases with the TPE corrections, it is not large enough to be
considered inconsistent with $P_2=0$.  In addition, the results from
Ref.~\cite{tvaskis06} are dominated by higher $Q^2$ points, where we
do not include nonlinearities in the TPE contributions from higher mass
intermediate states.  Including the single-experiment LT separations
from the new low $Q^2$ data sets used in this analysis, we find
$\langle P_2 \rangle = 2.8 \pm 2.4\%$ (after TPE), still generally consistent
with no nonlinearities.

\subsection{Extraction of $\ge$ and $\gm$ from global analysis}\label{sec:ltpt}

\begin{figure}[ht]
\includegraphics[height=12.0cm,width=8.4cm,angle=0]{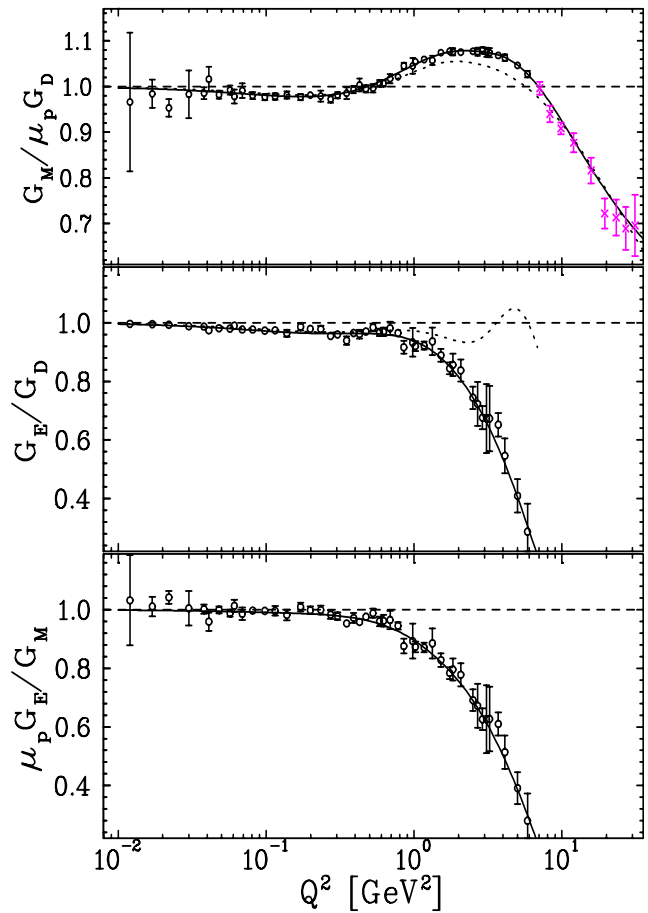}
\caption{(Color Online)
Extracted values of $\ge$ and $\gm$ from the global analyses.
The open circles are the results of the combined analysis of the cross section
data and polarization measurements (Sec.~\ref{sec:ltpt},
Tab.~\ref{tab:newffs}).  The magenta crosses are the extracted values of
$\gmp$ (Tab.~\ref{tab:newgm}) for the high $Q^2$ region, where $\gep$ cannot be
extracted.  The solid lines are the fits to TPE-corrected cross section and
polarization data (Sec.~\ref{sec:globfit}).  The dotted curves show the
results of taking $\ge$ and $\gm$ to be $\sqrt{\sigma_L}$ and
$\sqrt{\sigma_T}$, respectively, from a fit to the TPE-uncorrected reduced
cross section (Appendix~\ref{sec:param}), i.e. the value one would obtain
using only cross section data and ignoring TPE.
\label{fig:fitanddata}}
\end{figure}

In this section we extract individual $\ge$ and $\gm$ points and
uncertainties over the full $Q^2$ range where the form factors can
be separated.  The analysis follows that of the corrected cross section data
in the previous section, but now we include the PT measurements in each
$Q^2$ bin as part of the fit. The results for $Q^2 < 6$~GeV$^2$, where
$\gep$ and $\gmp$ can be separated, are given in Table~\ref{tab:newffs}, and
shown in Fig.~\ref{fig:fitanddata}.

For the combined fit, we fit $\ge$ and $\gm$ to the combination of cross
section and polarization transfer data in each small $Q^2$ bin.  For the
cross sections, we use the TPE-corrected cross section measurements, and
normalize each data set using the scale factors found in the global fit
(Sec.~\ref{sec:globfit}).  After making the initial fit for $\gep$ and $\gmp$,
we scale each data set by the estimated normalization uncertainty (as in
Sec.~\ref{sec:rosenbluth}) to find its contribution to the systematic
uncertainty, and add these in quadrature to determine the total uncertainty in
$\ge$, $\gm$ and the ratio due to the normalization uncertainties.

In addition to improving the overall precision, combining the cross section
and PT results has the added benefit of decreasing the correlation
in the uncertainties in $\ge$ and $\gm$.  The Rosenbluth separation tends
to yield a large anti-correlation between the uncertainties for $\ge$
and $\gm$, and thus an enhanced uncertainty on the ratio.  The PT data
measure the ratio directly, thus dramatically reducing this correlation.
Therefore, in Tab.~\ref{tab:newffs}, we provide values and uncertainties
for both the individual form factors and the form factor ratio.

\begin{table}[ht]
\caption{Extracted values for the corrected form factors relative to
the dipole form, and the ratio $\mugegm$.
\label{tab:newffs}}
\begin{ruledtabular}
\begin{tabular}{cccc}
$Q^2$[GeV$^2$] & $\ge/G_D$ & $\gm/(\mu_p G_D)$ & $\mugegm$  \\ \hline
  0.007 & 1.000$\pm$0.006 & 0.715$\pm$0.393 & 1.400$\pm$0.712 \\[-0.045cm]
  0.012 & 0.996$\pm$0.006 & 0.966$\pm$0.152 & 1.032$\pm$0.154 \\[-0.045cm]
  0.017 & 0.995$\pm$0.003 & 0.984$\pm$0.031 & 1.011$\pm$0.033 \\[-0.045cm]
  0.022 & 0.993$\pm$0.003 & 0.953$\pm$0.019 & 1.042$\pm$0.022 \\[-0.045cm]
  0.030 & 0.988$\pm$0.007 & 0.983$\pm$0.052 & 1.005$\pm$0.059 \\[-0.045cm]
  0.038 & 0.987$\pm$0.004 & 0.985$\pm$0.013 & 1.002$\pm$0.016 \\[-0.045cm]
  0.041 & 0.974$\pm$0.007 & 1.016$\pm$0.027 & 0.959$\pm$0.031 \\[-0.045cm]
  0.048 & 0.981$\pm$0.005 & 0.982$\pm$0.009 & 0.999$\pm$0.012 \\[-0.045cm]
  0.057 & 0.980$\pm$0.005 & 0.992$\pm$0.010 & 0.988$\pm$0.012 \\[-0.045cm]
  0.061 & 0.990$\pm$0.008 & 0.978$\pm$0.015 & 1.013$\pm$0.021 \\[-0.045cm]
  0.069 & 0.977$\pm$0.007 & 0.991$\pm$0.016 & 0.986$\pm$0.021 \\[-0.045cm]
  0.081 & 0.977$\pm$0.005 & 0.980$\pm$0.008 & 0.997$\pm$0.011 \\[-0.045cm]
  0.098 & 0.973$\pm$0.007 & 0.977$\pm$0.007 & 0.996$\pm$0.010 \\[-0.045cm]
  0.115 & 0.974$\pm$0.011 & 0.978$\pm$0.007 & 0.996$\pm$0.016 \\[-0.045cm]
  0.138 & 0.964$\pm$0.012 & 0.981$\pm$0.008 & 0.982$\pm$0.018 \\[-0.045cm]
  0.171 & 0.986$\pm$0.012 & 0.977$\pm$0.007 & 1.009$\pm$0.015 \\[-0.045cm]
  0.199 & 0.979$\pm$0.011 & 0.981$\pm$0.006 & 0.998$\pm$0.013 \\[-0.045cm]
  0.234 & 0.978$\pm$0.012 & 0.979$\pm$0.012 & 0.999$\pm$0.015 \\[-0.045cm]
  0.273 & 0.955$\pm$0.010 & 0.972$\pm$0.008 & 0.983$\pm$0.014 \\[-0.045cm]
  0.304 & 0.960$\pm$0.010 & 0.981$\pm$0.007 & 0.978$\pm$0.013 \\[-0.045cm]
  0.350 & 0.939$\pm$0.014 & 0.985$\pm$0.013 & 0.953$\pm$0.009 \\[-0.045cm]
  0.390 & 0.965$\pm$0.011 & 0.993$\pm$0.008 & 0.972$\pm$0.017 \\[-0.045cm]
  0.428 & 0.961$\pm$0.015 & 1.003$\pm$0.014 & 0.958$\pm$0.010 \\[-0.045cm]
  0.473 & 0.970$\pm$0.011 & 0.995$\pm$0.008 & 0.976$\pm$0.011 \\[-0.045cm]
  0.528 & 0.984$\pm$0.013 & 0.996$\pm$0.009 & 0.988$\pm$0.016 \\[-0.045cm]
  0.584 & 0.967$\pm$0.013 & 1.007$\pm$0.007 & 0.960$\pm$0.016 \\[-0.045cm]
  0.622 & 0.969$\pm$0.014 & 1.007$\pm$0.007 & 0.962$\pm$0.020 \\[-0.045cm]
  0.689 & 0.981$\pm$0.023 & 1.017$\pm$0.010 & 0.965$\pm$0.031 \\[-0.045cm]
  0.779 & 0.965$\pm$0.011 & 1.021$\pm$0.005 & 0.945$\pm$0.013 \\[-0.045cm]
  0.853 & 0.916$\pm$0.022 & 1.045$\pm$0.007 & 0.876$\pm$0.025 \\[-0.045cm]
  0.979 & 0.933$\pm$0.049 & 1.045$\pm$0.017 & 0.893$\pm$0.060 \\[-0.045cm]
  1.020 & 0.920$\pm$0.017 & 1.054$\pm$0.006 & 0.873$\pm$0.018 \\[-0.045cm]
  1.170 & 0.922$\pm$0.014 & 1.059$\pm$0.005 & 0.871$\pm$0.016 \\[-0.045cm]
  1.330 & 0.936$\pm$0.047 & 1.057$\pm$0.010 & 0.885$\pm$0.051 \\[-0.045cm]
  1.520 & 0.889$\pm$0.022 & 1.074$\pm$0.006 & 0.828$\pm$0.023 \\[-0.045cm]
  1.740 & 0.844$\pm$0.020 & 1.077$\pm$0.004 & 0.784$\pm$0.020 \\[-0.045cm]
  1.830 & 0.856$\pm$0.038 & 1.076$\pm$0.008 & 0.796$\pm$0.038 \\[-0.045cm]
  2.070 & 0.837$\pm$0.038 & 1.075$\pm$0.006 & 0.778$\pm$0.039 \\[-0.045cm]
  2.500 & 0.744$\pm$0.038 & 1.077$\pm$0.005 & 0.691$\pm$0.037 \\[-0.045cm]
  2.690 & 0.723$\pm$0.075 & 1.076$\pm$0.009 & 0.672$\pm$0.075 \\[-0.045cm]
  2.900 & 0.676$\pm$0.039 & 1.079$\pm$0.007 & 0.626$\pm$0.037 \\[-0.045cm]
  3.090 & 0.673$\pm$0.118 & 1.075$\pm$0.010 & 0.626$\pm$0.116 \\[-0.045cm]
  3.240 & 0.673$\pm$0.112 & 1.074$\pm$0.010 & 0.627$\pm$0.110 \\[-0.045cm]
  3.710 & 0.652$\pm$0.040 & 1.068$\pm$0.005 & 0.610$\pm$0.039 \\[-0.045cm]
  4.110 & 0.546$\pm$0.059 & 1.063$\pm$0.007 & 0.514$\pm$0.057 \\[-0.045cm]
  5.010 & 0.409$\pm$0.057 & 1.046$\pm$0.005 & 0.391$\pm$0.055 \\[-0.045cm]
  5.850 & 0.287$\pm$0.095 & 1.027$\pm$0.007 & 0.280$\pm$0.093 \\[-0.045cm]
\end{tabular}
\end{ruledtabular}
\end{table}

\subsection{Extraction of $\gm$ at high $Q^2$}\label{sec:gmp}

\begin{table}[ht]
\caption{Extracted values for the high $Q^2$ $\gmp$ results, along with the
$\gep$ values assumed in the extraction.
\label{tab:newgm}}
\begin{ruledtabular}
\begin{tabular}{ccc}
$Q^2$[GeV$^2$] & $\gm/(\mu_p G_D)$ & $\ge/G_D$ \\ \hline
   7.081 & 0.996$\pm$0.014 &  +0.054 \\[-0.045cm]
   8.294 & 0.940$\pm$0.018 & --0.058 \\[-0.045cm]
   9.848 & 0.908$\pm$0.013 & --0.191 \\[-0.045cm]
  11.990 & 0.877$\pm$0.021 & --0.364 \\[-0.045cm]
  15.720 & 0.816$\pm$0.028 & --0.629 \\[-0.045cm]
  19.470 & 0.721$\pm$0.033 & --0.814 \\[-0.045cm]
  23.240 & 0.713$\pm$0.039 & --1.062 \\[-0.045cm]
  26.990 & 0.689$\pm$0.047 & --1.273 \\[-0.045cm]
  31.200 & 0.696$\pm$0.067 & --1.564 \\[-0.045cm]
\end{tabular}
\end{ruledtabular}
\end{table}

In the extraction of $\gm$ for $Q^2 > 6$~GeV$^2$, the value of $\ge$
is not known, so that an additional assumption is required in order to
extract $\gm$.  We extract $\gm$ under two different assumptions for
the ratio $\gegm$.  First, we assume that the $\ge$ term is negligible
above 6~GeV$^2$, which would be the case if $\gep$ approached zero
and then stayed small.  Second, we assume a linear fall-off,
$\mugegm = 1 - 0.135\ (Q^2-0.24)$, from Ref.~\cite{gayou02}.
Up to $Q^2 \approx 14$~GeV$^2$ this yields a smaller contribution
from $\ge$ than in previous analyses, where it was assumed that
$\mugegm=1$.

At higher $Q^2$, the linear fit yields $|\mugegm | > 1$, and thus a
\textit{larger} $\ge$ contribution, almost 10 times what was
assumed in the inital analysis of the Sill, et al. data~\cite{sill93} at
$Q^2=30$~GeV$^2$.  This yields a significant change in the $Q^2$
dependence of the reduced cross section.  Instead of the electric
contribution decreasing from 6\% to almost zero as one went from
5 to 30~GeV$^2$, the linear decrease in $\ge$ implies almost no
contribution at 5--9~GeV$^2$, and a maximum contribution of
$\approx$ 6\% at the highest $Q^2$."

The final quoted values for $\gm$ are taken as the average of the values
obtained assuming $\mugegm=0$, and $\mugegm= 1 - 0.135(Q^2-0.24)$, with
half of the difference taken as the associated uncertainty.  Note that
the additional TPE component meant to estimate the high mass intermediate
state contributions, as discussed in Sec.~\ref{sec:uncertainty}, yields a
significant correction (and uncertainty) of 3--4.5\% to the cross section
over the range of the high $Q^2$ data.  Table~\ref{tab:newgm} shows the
extracted values of $\gmp$ based on the above assumptions, along with the
values of $\gep$ needed to reproduce the TPE-corrected cross sections when
combined with the quoted values of $\gmp$.  Because we average the extracted
$\gm$ values, and the cross section correction due to $\ge$ depends on
$\ge^2$, this corresponds to taking the average value of $\ge^2$ rather than
the average $\ge$.  Above $Q^2$=10~GeV$^2$, the only data included is from
Sill, et al.~\cite{sill93}.  There are also data from an earlier SLAC
experiment~\cite{coward67}, but both the scale and statistical uncertainties
are much larger than for the later measurements, and so these data are not
included.

\section{Conclusion}\label{sec:conclusion}

We have performed the first global analysis of elastic electron--proton  
scattering data taking into account two-photon exchange contributions and
their associated uncertainties.  The analysis combines the corrected
Rosenbluth cross section and polarization transfer data, and the  
corrected form factors $\ge$ and $\gm$ have been extracted over the
full range of $Q^2$ of existing data, up to $Q^2 \approx 6$~GeV$^2$
for $\ge$ and $Q^2 \approx 30$~GeV$^2$ for $\gm$.

The TPE corrections applied are based on the hadronic model of
Ref.~\cite{blunden05a} for the nucleon elastic intermediate state, but
with improved input form factors at the internal vertices.  These have
been supplemented by inelastic contributions estimated from recent   
calculations with explicit excited $N^*$ intermediate states
\cite{kondratyuk05, tjon07, kondratyuk07}, and from GPD-based partonic
calculations~\cite{chen04}.  The uncertainty in describing the
(off-shell) intermediate states is the main source of model dependence
in the analysis.  This uncertainty is more tractable, and relatively
mild, at low $Q^2$, but becomes larger at high $Q^2$ values.  For the
higher-mass intermediate state contributions, which are more important
for the high $Q^2$ data, we assign a 100\% uncertainty to the estimated
inelastic TPE correction.

The resulting TPE corrections to $\ge$ are significant in the region
$2 \ltorder Q^2 \ltorder 6$~GeV$^2$, where the LT and PT data are in
most striking disagreement, and here they bring the LT data into good
agreement with the PT results for $\gegm$. The corrections to $\gm$ are
smaller, but can be a few percent at moderately large $Q^2$ values.
We provide a convenient parameterization of the corrected $\ge$ and $\gm$ form
factors, as well as of the reduced elastic cross section, parameterized by
effective form factors which contain TPE effects.

While several recent quantitative studies have demonstrated that much of the
disagreement between the LT and PT results could be accounted for by TPE, none
of these calculations can be said to be complete, and none fully resolve the
discrepancy at all $Q^2$ values.  The hadronic calculations used here have the
least model dependence at lower $Q^2$, but require assumptions about the
additional high $Q^2$ corrections which have large uncertainties.  There
could in principle also be larger than expected TPE corrections from high mass
states to the PT experiments, although the indications from all existing
studies suggest that these are likely to have a small effect on the form
factor ratio. Additional measurements of TPE corrections will help to refine
these calculations by directly determining the TPE effects to both the cross
section and polarization measurements.

One such measurement is experiment E04-116 \cite{e04116}, using a beam of $e^+
e^-$ pairs produced from a secondary photon beam at Jefferson Lab, which will
make simultaneous measurements of $e^+ p$ and $e^- p$ elastic cross sections up
to $Q^2 \sim 2$~GeV$^2$.  The $e^+ p/e^- p$ ratio is directly sensitive to TPE
effects.  A proposal to perform a precise ($\sim 1\%$) comparison of $e^- p$ and
$e^+ p$ scattering at $Q^2=1.6$~GeV$^2$ and $\varepsilon \approx 0.4$ has also
been made at the VEPP-3 storage ring~\cite{vepp_proposal}.

With the inclusion of two-photon exchange, we also expect nonlinearities
in the $\varepsilon$ dependence of the cross section.  With the existing data,
the results do not show a significant nonlinearity before or after applying
TPE corrections, and so do not yet set significant limits.  A recently
completed Jefferson Lab experiment~\cite{e05017} will provide an accurate
measurement of the $\varepsilon$ dependence of the elastic $e p$ cross
section, with sufficient sensitivity to test the calculated nonlinearities.
Additional upcoming experiments~\cite{e04019,e04108} will extend polarization
transfer measurements to higher $Q^2$ values, and will examine in detail their
$\varepsilon$ dependence.  The increase in $Q^2$ will allow for a clean
extraction of $\ge$ and improved uncertainties on $\gm$ up to $\sim$9~GeV$^2$,
while the $\varepsilon$ dependence will be the first measurement sensitive to
TPE in polarization transfer.

On the theoretical front, the largest uncertainty arises from poor
knowledge of the contributions to the TPE amplitude from high mass
intermediate states, beyond the nucleon elastic contribution.  This
introduces significant model dependence in the correction, especially
at large $Q^2$.  It can be reduced by using phenomenological input
for the virtual Compton scattering amplitude at intermediate and high
$Q^2$ values.  In future it would be desirable to merge the low $Q^2$
hadronic calculations with the high $Q^2$ partonic approach, and
develop a framework which enables TPE corrections to be consistently
described over the entire $Q^2$ range of data.  The present analysis
should at the very least serve to highlight this need.  Finally, we may
expect in future more information on virtual Compton scattering to come
from lattice QCD, although calculation of four-point functions of this
type is currently still in its infancy.

\appendix
\section{Parameterization of Elastic Cross Section}\label{sec:param}

In this appendix we provide a convenient parameterization of elastic
cross section.  One can use the fits to $\ge$ and $\gm$ in
Sec.~\ref{sec:ltpt} to calculate the Born cross section, and then apply
radiative corrections, including TPE, to predict the measured cross
section.  However, this requires an explicit calculation of the TPE
correction to the cross section, consistent with the calculation used
to correct the cross section measurements.  It is useful for some
applications to work directly with the elastic cross section, without
necessarily requiring knowledge of the fundamental form factors.
We therefore provide a separate parameterization of the elastic cross
section and uncertainties {\em without} TPE corrections applied to the
data.  This will provide a simple and reliable, model-independent
parameterization for the cross section without direct reference to
TPE effects.

We parameterize the full elastic reduced cross section (Born + TPE
corrections) by:
\begin{equation}
\sigma_R^{\rm Born+TPE} =
\tau F_m^2(Q^2,\varepsilon) + \varepsilon F_e^2(Q^2,\varepsilon)\ ,
\label{eq:xsec_fit}
\end{equation}
where $F_m$ and $F_e$ are \textit{effective} magnetic and electric form
factors, respectively, which absorb the effects of multiple photon exchange.
Note that this is a purely phenomenological fit form, rather than a true
representation of the effect of two-photon exchange on the form factors,
which introduces a new term~\cite{guichon03}. In the Born approximation these
obviously approach the usual Sachs form factors:
\begin{equation}
F_m(Q^2,\varepsilon) \to \gm(Q^2)\ , ~~~~
F_e(Q^2,\varepsilon) \to \ge(Q^2)\ .
\label{eq:xsec_fit_born}
\end{equation}

In principle, the form factors $F_e$ and $F_m$ depend on both $Q^2$ and
$\varepsilon$, but as calculations and existing data~\cite{tvaskis06}
indicate extremely small deviations from linearity, we can at present
safely neglect this additional $\varepsilon$ dependence, and use the
form from Eq.~(\ref{eq:fitfcn}) when fitting $F_m$ and $F_e$.  We repeat
the global fit described in Sec.~\ref{sec:globfit}, using only
\textit{uncorrected} cross section data.  We keep the normalization
constants from the different data sets fixed to the values obtained
from the global fit, as these represent out best estimation of the
true normalization factors of the cross section measurements.  The
results of the fit for the cross section in Eq.~(\ref{eq:xsec_fit})
in terms of the effective form factors $F_e$ and $F_m$ are given in
Table~\ref{tab:xsec_fit}.

\begin{table}
\caption{Parameters for the fit to the TPE-\textit{uncorrected}
	cross section of Eq.~(\ref{eq:xsec_fit_born}), using the
	parameterization of Eq.~(\ref{eq:fitfcn}) for $F_m$ and $F_e$.
\label{tab:xsec_fit}}
\begin{ruledtabular}
\begin{tabular}{crr}
Parameter & $F_m/\mu_p$	& $F_e$~~ \\
\hline
 $a_1$  &  --2.151~~ &  --1.651~~ \\
 $a_2$  &    4.261~~ &    1.287~~ \\
 $a_3$  &    0.159~~ &  --0.185~~ \\ \hline
 $b_1$  &    8.647~~ &    9.531~~ \\
 $b_2$  &    0.001~~ &    0.591~~ \\
 $b_3$  &    5.245~~ &    0.000~~ \\
 $b_4$  &   82.817~~ &    0.000~~ \\
 $b_5$  &   14.191~~ &    4.994~~ \\
\end{tabular}
\end{ruledtabular}
\end{table}

\begin{acknowledgments}

We thank S.~Kondratyuk for providing the results from
Ref.~\cite{kondratyuk07}. This work was supported by the U. S. Department of
Energy, Office of Nuclear Physics, under contract DE-AC02-06CH11357 and
contract DE-AC05-06OR23177 under which Jefferson Science Associates, LLC
operates the Thomas Jefferson National Accelerator Facility.

\end{acknowledgments}

\bibliography{ffdata}

\end{document}